\documentstyle[mprocl]{article}

\input{psfig}
\def\Journal#1#2#3#4{{#1} {\bf #2}, #3 (#4)}

\def\EPL{\em Europhys. Lett.}
\def\IJMPB{{\em Int. J. Mod. Phys.} B}
\def\PRL{\em Phys. Rev. Lett.}
\def\PRB{{\em Phys. Rev.} B}

\def\JPC{{\em J. Phys.} C}
\begin{document}
\title{TWO-DIMENSIONAL DISORDERED ELECTRON SYSTEMS: A NETWORK MODEL APPROACH}
\author{P. FRECHE, M. JANSSEN, R. MERKT}
\address{Institute for Theoretical Physics, University of Cologne,
Z\"ulpicher Str. 77, \\ D-50937 K\"oln, Germany\\E-mail:
mj@thp.uni-koeln.de}
\maketitle \abstracts{We demonstrate that network models for wave
mechanical systems with quenched disorder cover the physics of
mesoscopic electrons. The models are constructed as a network of
random scattering matrices connecting incoming to outgoing wave
amplitudes. The corresponding wave dynamics is given by a discrete
unitary time evolution operator.  We report on three different
universality classes: two-dimensional, spinless, non-chiral electrons
with (O2NC) and without time reversal symmetry (U2NC), and
two-dimensional, non-chiral electrons with time reversal symmetric
spin-scattering (S2NC).  We determine the phase diagram in the
parameter space of scattering strengths.  The  O/U2NC models show strong
localization. We find symmetry factors in localization lengths as well
as multifractal exponents in agreement with theoretical predictions.
The S2NC model displays a localization-delocalization transition.  We
determine the critical exponent of the localization length and the
multifractal scaling exponent of the order parameter to be $\nu
\approx 2.4$ and $\alpha_0\approx 2.18$, respectively.}
\begin{figure}[b]
\centering \leavevmode \psfig{figure=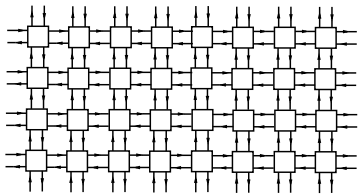,height=1.in}
\caption{A two-dimensional network of scatterers (squares) connecting
incoming and outgoing
propagating wave amplitudes (bonds).
\label{fig2}
}
\end{figure}
In phase coherent disordered electron systems (mesoscopic systems)
interference phenomena can lead to strong fluctuations in physical
quantities and localization of wave functions.
Localization-delocalization (LD) transitions tuned by system
parameters can be described in terms of the scaling theory \cite{Abr79}
for broad distribution functions of physical quantities \cite{JanR97}.
In the present work we start from a modeling of mesoscopic systems by
networks of unitary scattering matrices \cite{Sha82} (see
Fig~\ref{fig2}).  These models allow for a direct evaluation of
physical quantities and of (quasi-energy) eigenvalues and
eigenstates. Network models (NWMs) are paradigmatic for coherent waves
in disordered media and do not refer to any particular dispersion
relation. They are designed to cover essential symmetries and
characteristic length scales, but are otherwise unspecific. For
example, the wavelength  can be identified with the lattice
spacing which is, together with the wave velocity, set to unity. The
elastic mean free path $l_e$ is determined by the average reflection
properties of single scattering units.  NWMs can rather easily be
designed for special purposes. For example, the situation of strong
magnetic fields in 2D can be modeled \cite{ChaCod88,Fer88} by
suppression of forward and backward scattering in each scattering unit
of Fig.~\ref{fig2}.
\begin{figure}[t]
\centering \leavevmode
\psfig{figure=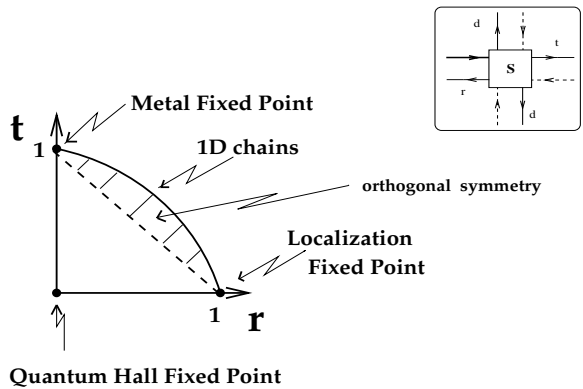,height=2.in,width=3.in}
\caption{The parameter space of the U2NC network model: $r^2 + t^2 \leq
1$ where $r$ and $t$ denote transmission strengths of individual
scatterers as shown in the inset. Vanishing deflection 
leads to decoupled 1D chains, maximum
transmission (reflection) corresponds to the metal (localization)
fixed point, and maximum deflection  corresponds to
the quantum Hall fixed point. Systems with time reversal symmetry (O2NC)
have a restricted parameter space, $r+t\geq 1$, denoted as `orthogonal
symmetry'.
\label{fig4}
}
\end{figure}
Consider a regular network of ${\cal N}$ sites and $N$ bonds as shown
in Fig.~\ref{fig2}.  The bonds carry propagating waves (incoming and
outgoing) represented by complex amplitudes. On the sites unitary
$S$-matrices map incoming to outgoing amplitudes.  The elements of
each $S$-matrix are (in general) random quantities taken from certain
distribution functions, characterized by a few parameters,
e.g. scattering strengths as shown in the inset of Fig.~\ref{fig4}.
Random phases (respecting symmetries) are attached to the amplitudes
on the links. They simulate random distances between scatterers in
realistic systems.  The construction of a NWM is fixed by the choice
of a type of $S$-matrix and a connectivity matrix $C$ that describes
how sites are connected to each other.  $S$ and $C$ define a unitary
time evolution operator $U$ \cite{Fer88,Edr88Zir94Kle95} that maps all
incoming to outgoing bond amplitudes in one unit of time,
\begin{equation}
	 U  \psi(t) = \psi(t+1) \label{1}\, .
\end{equation}
Here the state $\psi$ is the vector of the $N$ bond amplitudes.  The
eigenphases of $U$ are appropriate objects for investigating local
energy level statistics.  The NWM depicted in Fig.~\ref{fig2} (with
the scattering unit shown in the inset of Fig.~\ref{fig4}) is designed
to describe 2D disordered, non-interacting, spinless electrons in the
absence of chiral fields.  Its parameters are the transmission
(reflection) strength $t$ ($r$), and a deflection strength $d$ (equal
for left and right) obeying the constraint of unitarity,
$t^2+r^2+2d^2=1$. The parameter space is shown in Fig.~\ref{fig4}. The
model is denoted as U2NC model.  If the scattering is time reversal
symmetric the parameter space is further restricted by $r+t\geq 1$ and
the corresponding NWM is referred to as O2NC model.  Under real-space
renormalization the U2NC (O2NC) model has three (two) fixed points:
metal, localization, and quantum Hall fixed point \cite{ChaCod88}
(metal and localization fixed point).  We have calculated localization
lengths $\xi(M)$ in quasi-1D strip geometries of width $M$, and the
multifractal $f(\alpha)$ spectrum of eigenstates in a square geometry.
It turns out that only the localization fixed point is attractive
under renormalization while the others are repulsive, i.e.  all states
will localize for large enough system sizes.
\begin{figure}[t]
\centering\leavevmode \psfig{figure=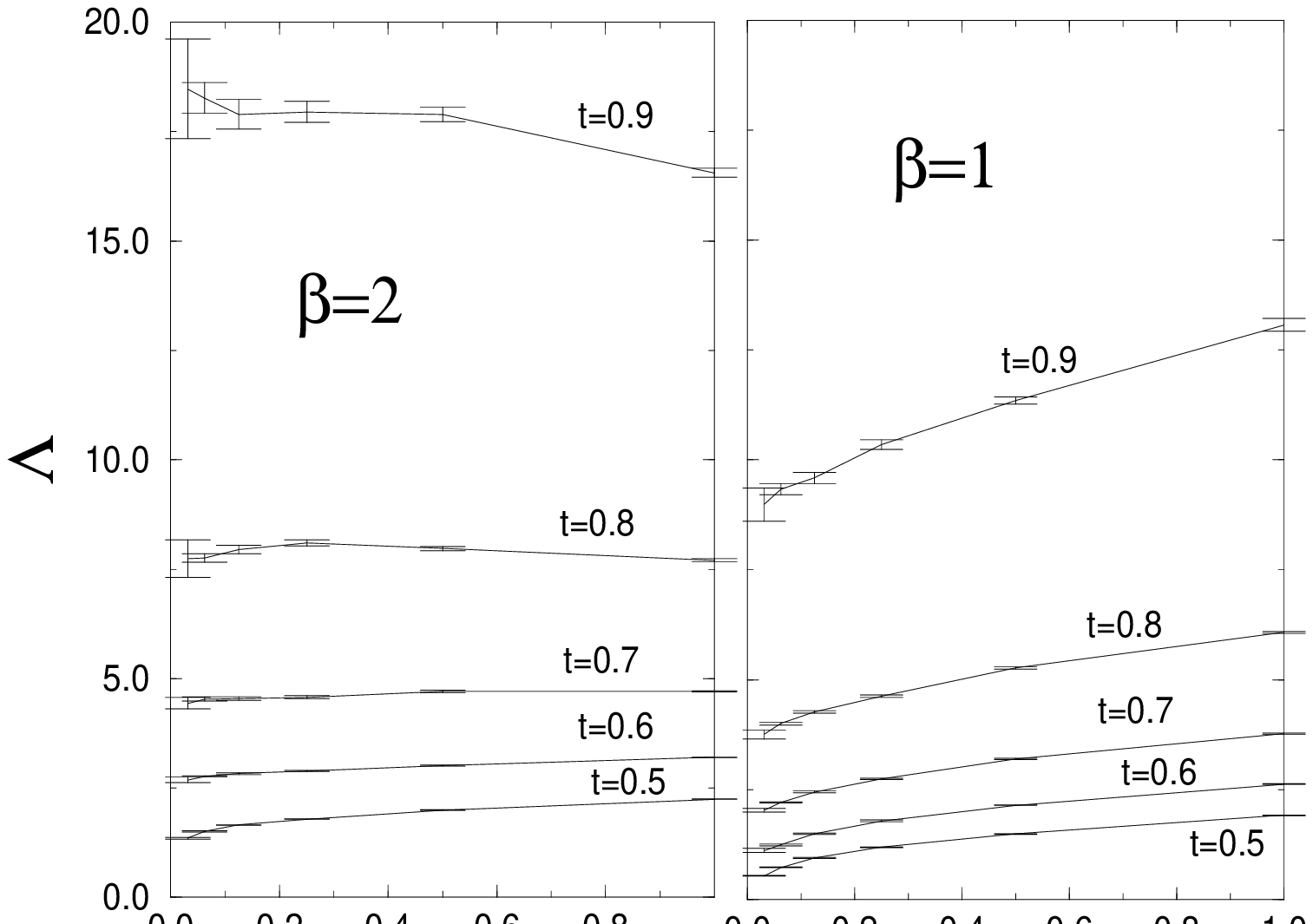,width=2.in}
\psfig{figure=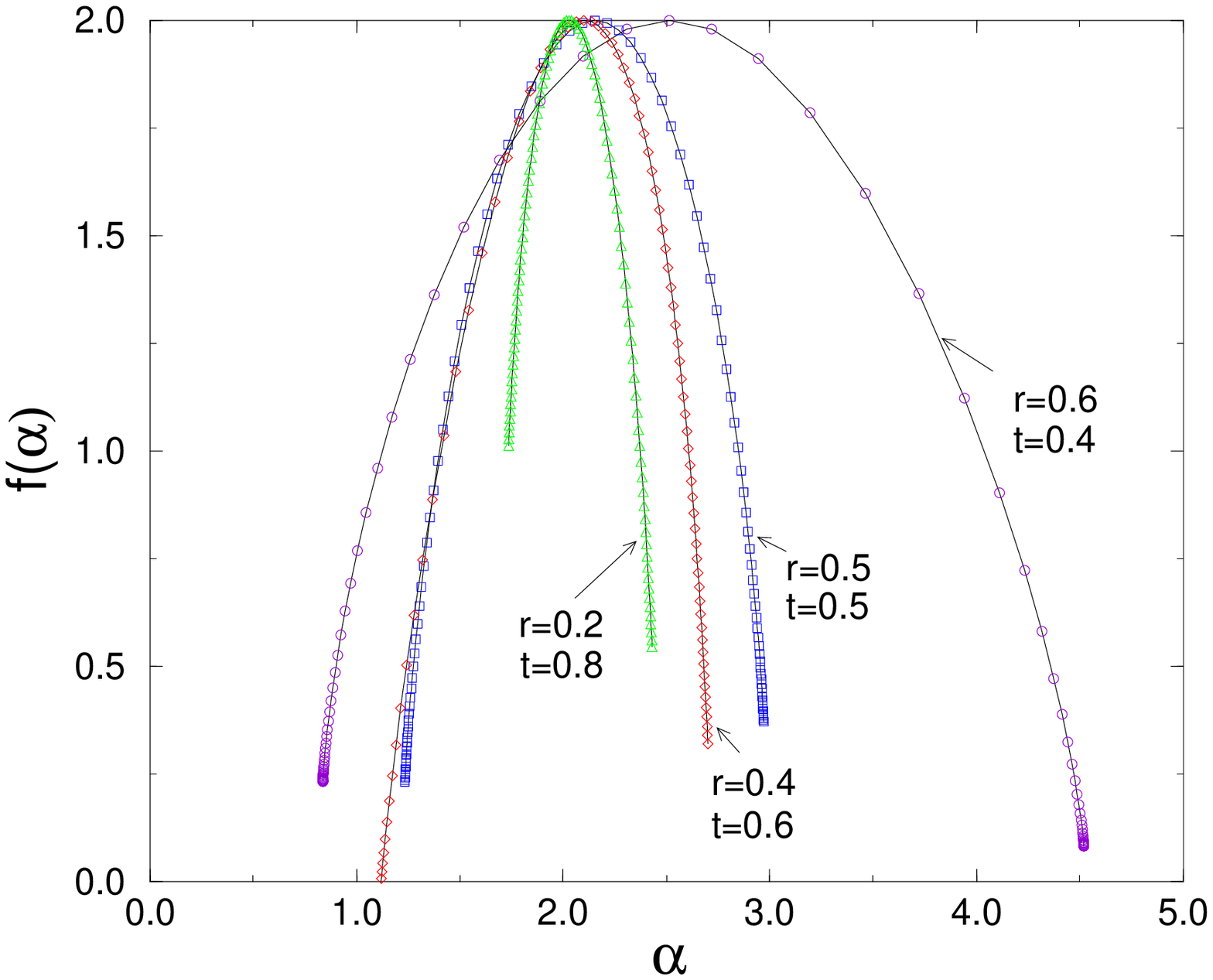,width=2.2in}
\caption{ On the left normalized localization lengths $\Lambda$ are
shown as function of the inverse strip width $1/M$ for different
symmetries ($\beta=1,2$) and different choices of the scattering
strengths $t$ ($r$ was set to $r=1-t$).  On the right
the multifractal spectra of states in a square geometry are shown for
different choices of the scattering strengths and $\beta=2$. The scaling
exponent $\alpha_0$ is given as the maximum position of $f(\alpha)$.
\label{fig5} }
\end{figure}
The elastic scattering length is given as
$l_e=  (t^2+d^2)/(2 r^2+2 d^2)$ (for details see
\cite{Fre97}) and by the Einstein relation a diffusive classical
conductance can be defined as $g_0=l_e/(2\pi)$.  As a quantitative
test we compared $\xi(M)$ with the analytic expression \cite{Cha92}
\begin{equation}
	\xi(M)=l_e(\beta (2M) +2 -\beta) \, .\label{3}
\end{equation}
Here $\beta=1,2$ is the usual symmetry index indicating presence ($1$)
or absence ($2$) of time reversal symmetry. Equation~(\ref{3}) is
expected to be valid in strip geometries for which $\Lambda=\xi(M)/M
\gg 1$. By finite size scaling this corresponds to a 2D metallic
system in the weak localization regime. In fact, we find good
agreement (with an uncertainty of $10\%$) already for moderate values
of $\Lambda {_> \atop ^\sim} 1$ (see Fig.~\ref{fig5}) telling that the
network model is able to describe disordered 2D electrons. As a second
quantitative test we calculated the scaling exponent $\alpha_0$
describing the scaling of the typical local density of states in a
square geometry \cite{JanR94}. Such states correspond to the weak
localization regime.  In 2D $\alpha_0$ can be related to $\Lambda(M)$
by a conformal mapping \cite{JanR94} as soon as $\Lambda$ becomes
independent of $M$ (in practice this means $\Lambda {_> \atop ^\sim}
3$ for $\beta=2$).
\begin{equation}
	\Lambda=\frac{1}{\pi(\alpha_0-2)} \longrightarrow \alpha_0-2=
 \frac{1}{2 \beta g_{0}}\label{4}
\end{equation}
This  finding coincides with the analytic result \cite{Fal95} 
obtained for large classical
conductance $g_0$ and is consistent with our numerical
results for $\alpha_0$, the maximum positions in the multifractal
$f(\alpha)$ spectra displayed in Fig.~\ref{fig5}.
\begin{figure}[t]
\centering
\leavevmode 
\psfig{figure=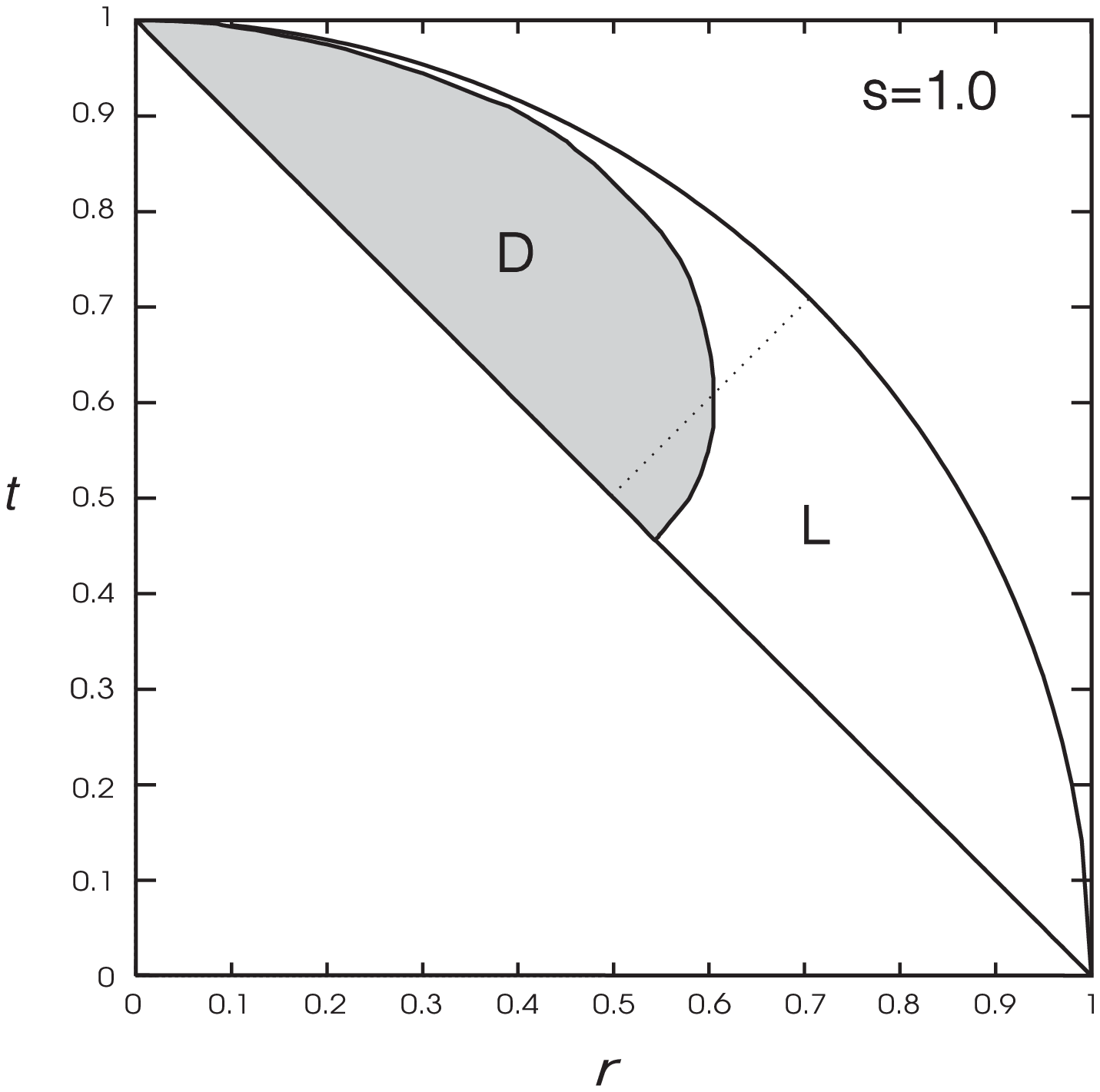,height=1.6in,width=1.6in}
\psfig{figure=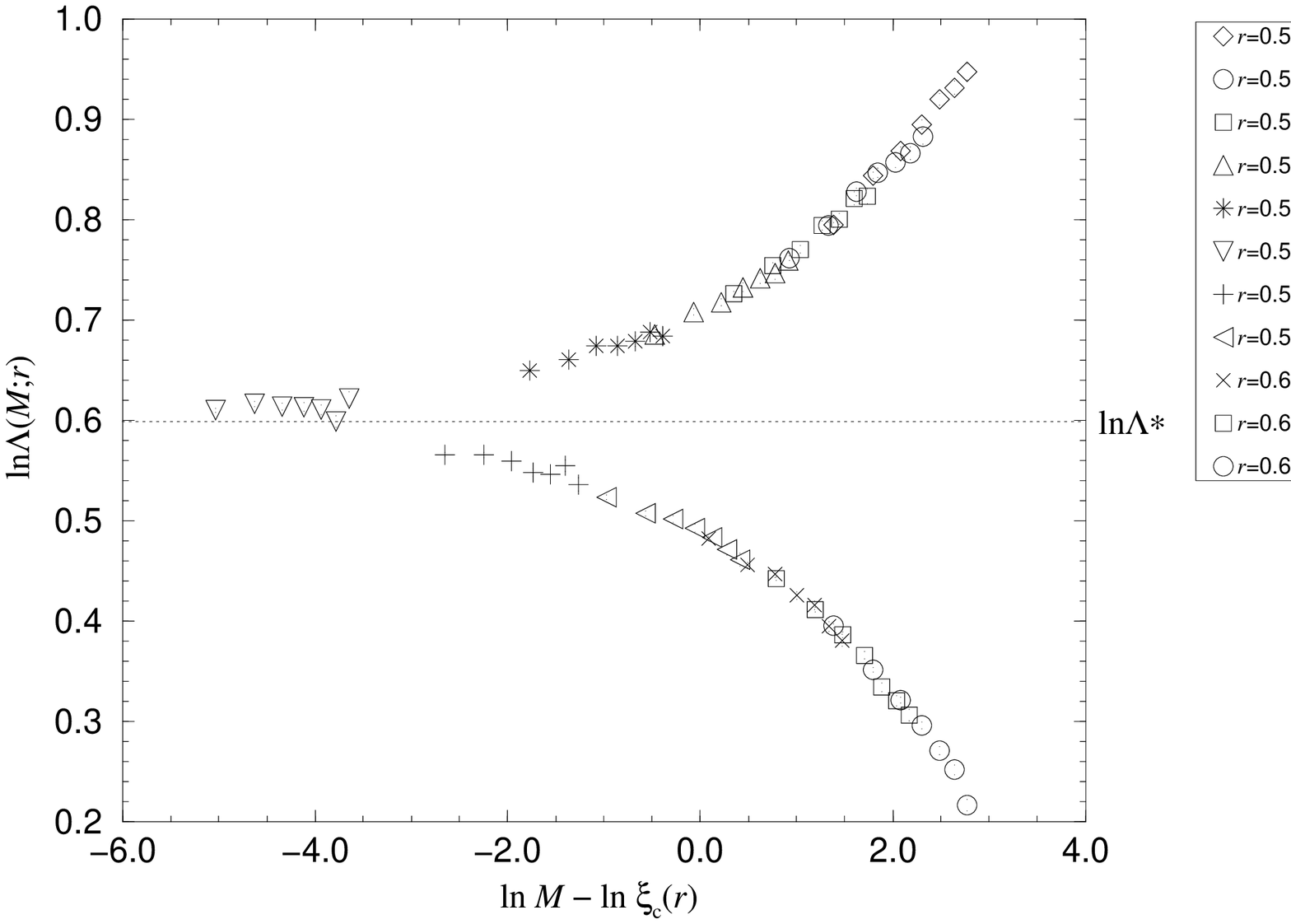,height=1.7in,width=1.7in}
\caption{ On the left the phase diagram of the S2NC model is shown for
maximum spin-scattering strength $s=1$. The grey area stands for the
delocalized phase (D) separated from the localized phase (L).  On the
right a one-parameter scaling function is shown for the LD transition
in the S2NC model. Data correspond to the logarithm of $\Lambda$ as a
function of $\ln \left(M/\xi_c(r)\right)$ where $\xi_c(r)$ denotes a
fitted correlation length as a function of the reflection strength
$r$.
\label{fig9}
}
\end{figure}
The NWM describing 2D disordered time reversal symmetric electrons in
the presence of spin scattering (S2NC model) is based on the O2NC
model \cite{Mer97}. In addition to the scattering parameters $t,r$ of
the O2NC model a new parameter, the spin scattering parameter $s\in
[0,1]$, appears. It defines a spin scattering length $l_{\rm S}(s)\in
[0,\infty[$.  The S2NC model shows true LD transitions in the
parameter space $(r,t)$ for non-zero values of $s$. The transitions
can be identified by finite size scaling techniques based on the
quantity $\Lambda(M)$.  The phase diagram obtained for maximum spin
scattering strength $s=1$ is shown in Fig.~\ref{fig9}. When crossing
the phase boundary in the $(r,t)$ plane, for a fixed value of $s$, the
quantity $\Lambda(M)$ follows one-parameter scaling \cite{Abr79} as
can be seen in Fig.~\ref{fig9}.  Analyzing the scaling function allows
for the determination of the critical exponent of the localization
length, $\nu \approx 2.4$ \cite{Mer97}.  With the help of the
conformal mapping relation (see Eq.~(\ref{4})) the multifractal
exponent can be obtained from the fixed point value of $\Lambda$ and
is $\alpha_0 \approx 2.18$.

In summary, we have shown that NWMs represent universality
classes of disordered wave mechanical systems. In particular, the U/O2NC
models show Anderson localization. A quantitative analysis in the
weak-localization regime is in reasonable agreement with known
analytic results for disordered electron systems.   The S2NC model
describing spin-scattering  exhibits localization-delocalization
transitions and allows for a quantitative analysis of critical
properties.

\eject


\section*{References}
\begin{thebibliography}{99}
\bibitem{Abr79}E. Abrahams {\it et al.}, \Journal{\PRL}{42}{673}{1979}.
\bibitem{JanR97}M. Janssen, preprint (cond-mat/9703196).
\bibitem{Sha82}B. Shapiro, \Journal{\PRL}{48}{823}{1982}.
\bibitem{ChaCod88}J. T. Chalker and P. D. Coddington,
\Journal{\JPC}{21}{2665}{1988}.
\bibitem{Fer88}H. A. Fertig, \Journal{\PRB}{38}{996}{1988}.
\bibitem{Edr88Zir94Kle95}I. Edrei {\it et al.},
\Journal{\PRL}{62}{2120}{1989}; M. R. Zirnbauer, \Journal{{\em Ann. 
Physik}}{3}{513}{1994}; 
R. Klesse and M. Metzler, \Journal{\EPL}{32}{229}{1995}.
\bibitem{Cha92}A. M. S. Macedo and J. T. Chalker,
\Journal{\PRB}{46}{14985}{1992}. 
\bibitem{Fre97}P. Freche, PhD thesis, University of Cologne (1997);
 to be published.
\bibitem{JanR94}M. Janssen, \Journal{\IJMPB}{8}{943}{1994}.
\bibitem{Fal95}V. I. Falko and K. B. Efetov, \Journal{\EPL}{32}{627}{1995}
\bibitem{Mer97}R. Merkt, Diploma thesis, University of Cologne (1997);
 to be published.


\end{thebibliography}
\end{document}